\magnification=\magstep1
\def\ltorder{\mathrel{\raise.3ex\hbox{$<$}\mkern-14mu
             \lower0.6ex\hbox{$\sim$}}}
\def\gtorder{\mathrel{\raise.3ex\hbox{$>$}\mkern-14mu
             \lower0.6ex\hbox{$\sim$}}}

\def\bF{{\bf F}}

\def\etal{{\sl et al.}\ }
\centerline{{\bf Basic $N$-Body Modelling of the Evolution of Globular
Clusters. I. Time Scaling}}
\bigskip
\centerline{S.J. Aarseth\footnote{$^1$}{Institute of Astronomy, 
Madingley Road, Cambridge CB3 0HA; e-mail\hfill\break
sverre@ast.cam.ac.uk} and D.C. Heggie\footnote{$^2$}{University of Edinburgh, Department of Mathematics and Statistics,
King's Buildings, Edinburgh EH9 3JZ; e-mail d.c.heggie@ed.ac.uk
}}

\vfill\eject

\noindent
Abstract
\medskip

We consider the use of $N$-body simulations for studying the evolution
of rich star clusters (i.e. globular clusters).  The dynamical processes
included in this study are restricted to gravitational (point-mass)
interactions, the steady tidal field of a galaxy, and instantaneous mass
loss resulting from stellar evolution.  With evolution driven by these
mechanisms, it is known that clusters fall roughly into two broad
classes: those that dissipate promptly in the tidal field, as a result
of mass loss, and those that survive long enough for their evolution to
become dominated by two-body relaxation.

The time scales of the processes we consider scale in different ways
with the number of stars in the simulation, and the main aim of the
paper is to suggest how the scaling of a simulation 
should be done so that the results
are representative of the evolution of a ``real''
cluster.  We investigate three different ways of scaling time.  One of
these is appropriate to the first type of cluster, i.e. those that
dissipate rapidly, and similarly a second scaling is appropriate only to
the second (relaxation-dominated) type.  We also develop a hybrid
scaling which is a satisfactory compromise for both types of cluster.
Finally we present evidence that the widely used Fokker-Planck method
produces models which are in good agreement with $N$-body models of
those clusters which are relaxation dominated, at least for $N$-body
models with several thousand particles, but that the
Fokker-Planck models evolve too fast for clusters which dissipate
promptly.

\vfill\eject

\bigskip
\noindent
1. Introduction
\medskip
This paper is concerned with techniques for studying the dynamical
evolution of a globular star cluster.  This is a complicated problem:
a complete description would be impossible without a detailed
understanding of the way in which stars interact non-gravitationally,
and how the internal evolution of the stars interacts
with their dynamical behaviour.  This is illustrated by the emerging
study of the behaviour of binaries in star clusters (Milone and Mermilliod
1996).  Nevertheless there is no astrophysical problem which admits of a
complete description, and much may be learned by the study of more
tractable, idealised models, provided  that the goal of
understanding the behaviour of real clusters is always kept in mind.

To a very good approximation the internal dynamics of a globular star
cluster is an $N$-body problem.  The dominant forces which govern the
internal motions (i.e. motions relative to the centre of mass of a
cluster) are the gravitational forces of the stars themselves and the
tidal field of the parent galaxy.  In addition, the masses of the stars
are constant for long periods of time.  Nevertheless, stars lose significant
amounts of mass from about the time at which they complete their main
sequence evolution, and it is known from simplified models
(e.g. Applegate 1986, Chernoff \& Shapiro 1987)  that the effects on the cluster can be
devastating.  It is fair to say that this is the principal known channel
through which the internal evolution of the stars can influence the
gross dynamics of a cluster.

We have now stated the three main processes whose effects we explore
in this paper:  (i) mutual gravitational interactions of the stars;
(ii) the tidal field of the galaxy; (iii) mass loss by stellar
evolution.  Clearly much has been omitted.  For example, nothing has
been said about the effects of primordial gas, whose early dissipation
may well be sufficient to unbind a young cluster (e.g. Lada \etal
1984, Goodwin 1997).  Nevertheless, as such studies show, the
influence of this matter may well almost cease very early on in the
evolution of a cluster, i.e. by the time the most massive stars have
evolved. Another omission is the influence of primordial binaries
(e.g. Hut \etal 1992).   Their effect is more persistent, and indeed
they may well be crucial for long periods in the later evolution of a
cluster, probably determining such factors as the size of the core
(Goodman \& Hut 1989, Gao \etal 1991, Vesperini \& Chernoff 1994).
Nevertheless, up to the point of core collapse (the time scale for
which is one of the features on which we concentrate in this paper),
primordial binaries are thought to behave roughly as a species of
rather massive stars, i.e. with a mass equal to the sum of the masses
of the components (Heggie \& Aarseth 1992).  Up to this point, then,
it is not expected that they have a very important effect on the gross
dynamics.

The simplified problem that we have stated can be tackled by more than
one technique.  At present the most popular is the Fokker-Planck model,
which treats a cluster of stars rather like a gas, the gravitational
encounters between the stars being modelled in much the same way as
collisions in an ideal gas.  This technique is idealised in a number of
ways: often it is assumed (i) that the cluster and tidal field are
spherically symmetric, (ii) that the distribution of stellar velocities is
isotropic, (iii) that the evolutionary time scale is long compared with the
time scale on which stars orbit within the cluster, and (iv) that small-angle
scattering encounters are dominant.  Techniques for dispensing with 
most of
these assumptions are known (e.g. (i) Goodman 1983a, Einsel \& Spurzem 1998; 
(ii) Takahashi
1995; (iii) Spitzer \& Hart 1971; (iv) Goodman 1983b), but they are not in
routine use at present.

Apart from Fokker-Planck techniques, another potential contender is
the $N$-body method (e.g. Aarseth 1985).  This has no difficulty with
any of issues (i)-(iv) raised above.  Indeed there are important ways in
which $N$-body techniques become easier if the models are made more
realistic.  One example is the effect of introducing a spectrum of
stellar masses.  Multi-mass Fokker-Planck models are more time-consuming
than models in which all stars have the same mass, whereas $N$-body
models are easier because the time scale for evolution becomes 
shorter.

In general $N$-body models are more free of simplifying assumptions than
other techniques.  Their great drawback is that it is still impractical
to model systems with the correct $N$, because of the computing time
required.  On a general-purpose supercomputer a 10,000-body model
requires of order 2 months to pass core collapse (with equal masses;
Spurzem \& Aarseth 1996), while even the fastest special-purpose
computer (the GRAPE-4 in Tokyo) takes of order 3 months for about 32,000
particles (Makino 1996; the calculation was actually performed with only
one quarter of the full configuration of the hardware).  These are small
simulations compared with a median globular cluster, which may contain
of order $10^6$ stars.  Therefore the main problem in studying the
evolution of globular clusters with $N$-body models is knowing how the
results depend on $N$.  

To some extent this question could be answered empirically, by computing
simulations with different $N$ and extrapolating.  Theory, however,
gives important information on the $N$-dependence of known evolutionary
processes such as two-body relaxation, and it is desirable that the
extrapolation should be consistent with this.  It follows that
the application of $N$-body techniques to the
evolution of globular clusters is to some extent as dependent on theory
as the Fokker-Planck model is.

In this paper we have two aims.  One aim is to show that the results of
$N$-body simulations can indeed be scaled reliably to systems of the
size of globular clusters.  We do so primarily by comparing simulations
with different $N$, the scaling being determined in accordance with
theoretical considerations.  Our second aim, however, is to demonstrate
how the resulting evolution compares with that obtained with
Fokker-Planck techniques incorporating the same physical mechanisms.  

To some extent both questions have already been studied by Fukushige and
Heggie (1995), who compared the results of $N$-body and Fokker-Planck
calculations, and varied the number of particles in the simulations
until the results were nearly independent of $N$.  That paper imposed on
itself two limitations, however.  First, the paper was concerned only
with clusters in which relaxation effects were unimportant:  the models
dissolved in the tidal field before there was any significant mass
segregation. The second limitation, which derives from the first, was
that the scaling of the models to real star clusters was uncomplicated,
essentially because it was unnecessary to scale the results so that
relaxation effects in the models would closely mimic those in a real
cluster.  

In the present paper we remove these limitations, and extend 
the problem by considering how to model 
those clusters in which relaxation is
important, which in effect means those in which there is relatively
little mass loss by stellar evolution, and/or the initial configuration
is sufficiently centrally concentrated that relaxation effects can occur
sufficiently quickly.  

Here is the plan of the paper.  In the next section we introduce the
physical mechanisms which drive the evolution, summarise our choice of
initial conditions, outline some of the hardware and software aspects
of our simulations, and discuss the central issue of time scaling.  
Section three describes the results, which show
that clusters whose evolution is driven by tidal overflow, two-body
relaxation, and mass-loss from stellar evolution, may indeed 
be simulated with
relatively modest values of $N$.  We also show that the results of
simulations with a few thousand particles are
remarkably consistent with those already found using Fokker-Planck
techniques, at least for those clusters which are not so seriously
affected by the third mechanism (i.e. mass loss by stellar evolution) 
that they disrupt in a time much less
than a Hubble time.  In the last section of the paper we summarise our
conclusions, and also draw attention to some astrophysically
interesting features of the evolution, such as the anisotropy of
velocities.  

A companion paper (Vesperini \& Heggie 1997) applies some of the
techniques of the present paper to discuss the evolution of the mass
function.  Finally, the question of how to scale $N$-body models so that
the statistics of stellar collisions in evolving globular clusters may
be studied is the subject of a further paper (Aarseth \& Heggie 1998).

\bigskip
\noindent
2. Technique
\medskip\noindent
2.1 Initial and Boundary Conditions
\smallskip
As has been stated, one of our aims has been to compare the results of 
$N$-body simulations
with those of Fokker-Planck calculations.  Fortunately there exists an
excellent and instructive set of such calculations described by Chernoff
\& Weinberg (1990), and so we have adopted their initial and boundary
conditions (with appropriate modifications described below), even where
these differ from those that might now be considered more appropriate
for the study of galactic globular clusters (e.g. the minimum mass).  In
one or two cases we have carried out a few $N$-body simulations to
observe the effect of changing these parameters, but much more detailed
results along these lines will be found in the paper by
Vesperini \& Heggie (1997).

Chernoff \& Weinberg studied models of clusters in circular orbit about a galaxy
whose gravitational field is taken to be that of a point mass.  The
circular speed is $v_g = 220$km/s.   In our $N$-body simulations we
have adopted the corresponding equations
$$\eqalign{
\ddot x - 2\omega\dot y - 3\omega^2x &= F_x\cr
\ddot y + 2\omega\dot x&= F_y\cr
\ddot z + \omega^2z &= F_z\cr
}\eqno(1)
$$
(e.g. Chandrasekhar 1942), where the coordinates of a star are taken
relative to a rotating frame whose origin rotates about the galaxy in a
circular orbit of angular velocity $\omega$; the $x$- and $y$-axes point in the
direction away from the galactic centre and in the direction of galactic
rotation, respectively; and $\bF$ is the acceleration due to other
cluster members.  Note that Fokker-Planck codes usually impose a cutoff,
rather than a tidal field as in the $N$-body simulations, though the
evidence is that this makes little difference to the rate at which mass
is lost (Giersz \& Heggie 1997).

For initial conditions Chernoff \& Weinberg
(1990) used King models (King 1966) with
scaled central potential $W_0 = 1$, $3$ and $7$.  These models have also
been used in the present study, but with $W_0 = 3$, $5$ and $7$. The
reason for ignoring the models of lowest concentration 
is that Chernoff \& Weinberg 
found that they always disrupted in much less than a
Hubble time, a result that has already been confirmed (with substantial
quantitative differences) with $N$-body models by Fukushige \& Heggie
(1995).  Chernoff \& Weinberg found that, depending on the initial mass
function, clusters with $W_0 = 3$ and $7$ could survive until the present day,
and we have added the case $W_0 = 5$ to help reveal trends in the
results.  In planning our programme of work we regarded the results of
Fukushige \& Heggie as providing a definitive $N$-body answer to the
evolution of those clusters which dissolve quickly, without surviving
long enough for the effects of relaxation to become important. 
Therefore we have concentrated on those models whose
long-term fate is determined in an important way by relaxation effects,
and was left open by Fukushige \& Heggie.

We have followed Chernoff \& Weinberg in the choice of initial mass
function, which is a power law $dN\propto m^{-\alpha}dm$ in the range
$0.4M_\odot<m<15M_\odot$, with $\alpha = 1.5$, $2.5$ or $3.5$.  We have
also followed their prescription for mass loss through stellar
evolution, which is assumed to take place instantaneously at the end of
main sequence evolution.  
The time and amount of mass lost are
determined by linear interpolation in a table.  At the time of its
formation a remnant is assumed to have exactly the same velocity as its
progenitor;  in making this assumption we are again following 
Chernoff \& Weinberg, though it is likely to be in serious error for 
neutron stars, only a small fraction of 
which would probably be retained
(Drukier 1996).  There are no primordial binary stars in our models,
though these will be included in the next paper in this series
(Aarseth \& Heggie 1998).

Let us consider next the number of parameters in this problem.  There
are two dimensionless parameters, $W_0$ and $\alpha$, which we retain, and two
dimensional parameters, which are the initial mass of the cluster,
$M(0)$, and its galactocentric distance, $R_g$.  As shown by Chernoff \&
Weinberg, clusters with the same relaxation time exhibit the same
evolution, if the assumptions of the orbit-averaged Fokker-Planck model
are valid.  Therefore the evolution should depend on $M(0)$ and $R_g$
only through the parameter $F$ introduced by Chernoff \& Weinberg, as
this is a measure of the relaxation time.  We follow Chernoff \&
Weinberg in presenting results for four distinct values of $F$,
corresponding to the four ``families" that they studied.
In Sec.3.2 we check whether different models within the same family
evolve similarly.

\medskip\noindent
2.2 Practical considerations
\smallskip

In carrying out our $N$-body models we have employed a version of the
code NBODY4 (Aarseth 1985), which may be briefly described as follows.  
It is a direct summation code with a Hermite integrator (Makino  \& Aarseth
1992)  and a binary hierarchy of time-steps (i.e. ``block" time
steps).  It employs various mechanisms for the treatment of compact
subsystems, including two-body and chain regularizations (Mikkola \& Aarseth
1993).  The inner binary of a hierarchical triple system is treated by a
novel technique for rescaling of time (Mikkola \& Aarseth 1996).  The main
feature of other advanced codes which NBODY4 lacks is the distinction
between irregular and regular forces (Ahmad \& Cohen 1973), i.e. those
due to near neighbours and those due to the remainder of the system.

The evaluation of forces and force derivatives (which are also required
in the Hermite integrator) has been carried out with the aid of a
special-purpose computer.  This is a single GRAPE-4 board containing 48
HARP 
processors (cf. Makino \etal 1993), which also performs one or two other
compute-intensive tasks, such as finding neighbours and potentials.  The
board connects to a standard Dec Alphastation 3000/700 via a HARP
control board and interface.  The performance of the board is summarised
in Table 1; most of the results in this table are discussed in \S3
below, but it also gives the CPU time for a variety of models.  It can be
seen that most models take less than one day, which is a reasonable target
for a programme in which many models must be computed in order to
explore an adequate range of parameters.  

It is interesting to compare
the timing with that of the full-size GRAPE-4 in Tokyo (Makino \etal
1997).  They timed equal-mass King models with $W_0 = 3$ using a single
cluster of nine processor boards, and we assume that our models with the
same $W_0$ and the 
steepest mass function ($\alpha = 3.5$) are the most comparable.
We
find that our single board (with associated host) takes about $0.97$
minutes for a single $N$-body time unit (Heggie \& Mathieu 1986) 
if $N = 4096$, and this 
appears to be only about twice as long as on the nine boards in a single
cluster in Tokyo.  For $N = 8192$ the corresponding time for our
single board is about
$2.42$ minutes, which exceeds the result of Makino \etal (on a
cluster of nine boards) by a factor of a little more than two.  In fact our timings
are quite insensitive to the slope of the mass function.  

The fact that our system performs at almost half the speed of one whose
peak speed is nine times higher can be interpreted in terms of
efficiency, i.e the fraction of peak speed actually achieved.  As shown
by Makino \etal (their Fig.17) the efficiency of their hardware is about
$8$\% when $N = 8192$ (for a certain model).  It may be expected that a
smaller system, though slower, would be more efficient.  Indeed, we
estimate the efficiency of our hardware at about 25\% for the same $N$,
and this largely accounts for the relative timings.

Now we turn to some practical aspects in the setting up of the initial
$N$-body model.  Using the units introduced by King (1966), the desired
King model was first constructed by numerical integration of Poisson's equation,
which also
yields the corresponding value of the tidal radius $r_t$.  Then an
$N$-body realisation of the model was constructed, also in King's units.
This $N$-body model was then scaled to standard $N$-body units (Heggie
\& Mathieu 1986), in which the total mass of the cluster is $1$, the
constant of gravitation is $1$ and the initial energy is $-1/4$.  {\sl
Henceforth quantities expressed in these units will be distinguished by
$\ast$.}  The scaling was carried out so that this $N$-body model,
regarded as an isolated model, is in virial equilibrium.  The
theoretical tidal radius of the model in King units was scaled in the
same way, giving a value $r_t^\ast$ for the tidal radius of the $N$-body
model in $N$-body units.  Now in these units it follows from eqs.(1)
that the theoretical tidal radius is also given by 
$$ r_t^\ast =
(3\omega^{\ast2})^{-1/3},\eqno(2) $$ 
and so the value of $\omega^\ast$
is determined.  Note that the resulting initial 
model is not quite in virial
equilibrium, because of the contribution to virial balance of the terms
involving $\omega$ in eqs.(1) (cf. Chandrasekhar 1942).  This could be
rectified (Heggie \& Ramamani 1995), but the neglected 
contribution is no more than a few percent, and is not noticeable
in comparison with statistical fluctuations in the virial ratio.

\vfill\eject
\noindent 
2.3 Scaling 
\smallskip 

We now consider how the model so constructed is to be scaled in such a
way that it represents the evolution of a ``real" star cluster. As we
shall see, the most delicate aspect of this question is how to scale the
time from the $N$-body units of the simulation to some astrophysical
unit of time such as $10^6$yr.  One of the principal roles of this
scaling is that it determines the points in the simulation at which mass
loss by stellar evolution takes place.  In this subsection we first
state two scalings which have an obvious physical basis, and then 
present a somewhat novel hybrid, which
is designed to combine their advantages.

\smallskip\noindent
2.3.1 Fixed Scalings
\smallskip

The simplest choice is scaling by the crossing time, i.e. we write
$$
t = t^\ast {t_{cr}\over t_{cr}^\ast},\eqno(3)
$$
where $t_{cr}$ denotes the crossing time of the real cluster (in Myr,
for example), and $t_{cr}^\ast$ denotes that of the $N$-body model (in
$N$-body units).  This choice guarantees that the period of radial
motion of a particle in the simulation scales to the period of radial
motion of a star in the cluster.  
It may be shown readily that it also 
ensures that the angular velocity $\omega^\ast$, defined by eq.(2),
scales correctly to the orbital angular velocity of the cluster, i.e. to
$\omega = v_g/R_g$.  Thus eq.(3) is equivalent to assuming that $t =
t^\ast U_t$, where
$$
U_t = \omega^\ast/\omega.\eqno(4)
$$
In general, scaling by the crossing time is appropriate at times 
when mass loss is important, because simple arguments show 
(Hills 1980)
that the effect  of {\sl impulsive} mass loss, i.e. mass loss taking
place on a time scale much smaller than the crossing time, is
quantitatively very different from that of adiabatic mass loss.

Scaling by the crossing time, eq.(3), does not
ensure that the number of relaxation times that elapse in the
two systems correspond correctly, because the relaxation time is an
$N$-dependent multiple of the crossing time, and the number of particles
in a practical simulation is much smaller than that in the globular
clusters of interest. A modified time
scaling, which ensures that {\sl relaxation} proceeds at the correct rate, is 
$$ 
t = {t_{rh}\over t_{rh}^\ast}t^\ast,\eqno(5) 
$$ 
where $t_{rh}$ is the half-mass relaxation time in the real cluster, and
$t_{rh}^\ast$ is the corresponding quantity for the $N$-body model in
$N$-body units.  (Any other measure of relaxation times, such as the
central value, would be equally suitable.)

\smallskip\noindent
2.3.2 Variable Scaling
\smallskip

Scaling by the crossing time, i.e. eq.(3), was used by Fukushige \& Heggie (1995) to
simulate the early evolution of globular clusters, but this meant that
these authors were unable to determine the long-term fate of the
clusters which survived this early phase, i.e. those with steep initial
mass functions and/or high initial concentration, for the evolution of
such clusters becomes relaxation-dominated. It is important to be able
to simulate such clusters, for if the initial conditions of the galactic
globular clusters bear any resemblance to those of the models surveyed
by Chernoff \& Weinberg, it is these long-lived models which are
relevant to the  evolutionary history  of the clusters which exist
today.   Ideally, therefore, one should devise a strategy which combines
the advantages of both kinds of scaling, i.e. eq.(3) for a cluster which
dissolves promptly, and eq.(5) for a long-lived cluster, at least 
during the long phase of its evolution which is dominated by relaxation.
In order to decide which scaling to adopt, and when,  we may compare
the time scales for the two dominant processes, i.e. mass-loss and
relaxation.

The time scale for mass-loss by stellar evolution may be defined as 
$$
t_M
= - M/(dM/dt),\eqno(6)
$$ 
where the derivative on the right indicates only mass
loss arising from this mechanism (cf. eq.[8] and Fig.1).  
We denote the time scale of
two-body relaxation by $t_r$, and defer a precise definition until later
in this section.  Then our strategy suggests that we adopt eq.(5), i.e.
scaling by the relaxation time, if $t_M>t_r$.  We must be more careful, however,
before concluding that we use eq.(3) if $t_M<t_r$, because this scaling
does not correctly scale the relaxation time of the cluster to that of
the model.  Therefore a situation could arise in which $t_r^\ast <
t_M^\ast$, even though $t_r > t_M$.
For
this reason we introduce an intermediate regime in which we compromise
as best we can, by adopting a
scaling which ensures that $t_r^\ast = t_M^\ast$.  It provides a smooth
transition from the scaling of eq.(3) to that of eq.(5) as the mass loss
from stellar evolution gradually slows down.

At this point we summarise the {\sl variable} scaling that is adopted in
the bulk of the simulations described in this paper.  
Because we must switch between one scaling and another we will not
usually be able to express $t$ as a simple multiple of $t^\ast$, and
instead must consider scalings in differential form. Thus we have
$$
{dt\over dt^\ast} = \cases{\displaystyle{{t_{cr}\over t_{cr\ast}}} &if
$t_M\le t_{cr}\displaystyle{{t_r^\ast\over t_{cr}^\ast}}$,\cr
\displaystyle{{t_{M}\over t_{r}^\ast}} &if
$t_{cr}\displaystyle{{t_r^\ast\over t_{cr}^\ast}}\le
t_M\le t_r$,\cr
\displaystyle{{t_{r}\over t_{r}^\ast}} &if 
$t_M\ge t_{r}$.\cr}\eqno(7)
$$

Now we must consider how this may be implemented in an $N$-body
calculation.  We assume that, at any time, 
there is a power-law mass function with
$N(m)dm\propto m^{-\alpha}dm$ for $m_1>m>m_0$.  Then the total mass is
$M\propto m_1^{2-\alpha} - m_0^{2-\alpha}$ and if we assume (for this
purpose only) that each star loses {\sl all} its mass at the end of its
evolution, then it is easy to show that 
$$ 
t_M =
{\left(1-\left(\displaystyle{{m_0\over m_1}}\right)^{2-\alpha}\right)
t\over (\alpha-2)\displaystyle{{d\log m_1\over d\log t}}}.\eqno(8) 
$$
This is easily computed if the current value of $m_1$ is known (i.e. the
mass of a star which is just reaching the end of its evolution) and if
we assume that $\alpha$ retains its initial value; and the
value $\displaystyle{d\log m_1\over d\log t}$ is easily computed from
the table of lifetime versus mass. Fig.1 shows the result for several
values of $\alpha$.

In eq.(7) the value of $t_M$ must be compared with $t_{r}$, and we must
adopt a suitable definition of the relaxation time which allows us to
determine whether the evolution of a cluster is dominated by relaxation
or mass-loss.  This is not straightforward.  The evolution of the core
of a high-concentration cluster may be dominated by relaxation even
though mass-loss causes the half-mass radius to expand.  The issue is
further complicated by the mass-dependence of mass segregation, which
depends in turn on the evolving mass function.  Despite these
complications we have adopted a simplistic criterion based on a result
of Spitzer (1987, his eq.[4-1]) for values at the half-mass radius.  We
write it in the form 
$$
{t_r\over t_{cr}} = {N\over 11\ln(\gamma N)}.\eqno(9)
$$
The current value of $N$ may be obtained from the current value of
$N^\ast$ and the initial values $N(0)$
and $N^\ast(0)$, where $N(0) = M(0)/\bar m(0)$, if we assume that
$N/N^\ast$ is constant, and this is correct provided that the model
correctly simulates the evolution of the real cluster.

The remaining quantities in eq.(7) are easily calculated.  Thus
$
t_{cr} = U_t t_{cr}^\ast,
$  
where $U_t$ is given by eq.(4), and $t_{cr}^\ast$ may be computed from
the conventional definition 
$t_{cr}^\ast = M^{\ast5/2}/(2\vert
E^\ast\vert)^{3/2}$, in which $E^\ast$ is the energy of the model in
$N$-body units.   The ratio $t_r^\ast/t_{cr}^\ast$ is obtained in
analogy with eq.(9).

We close with a few practical details.  First, we adopt a relatively low
value of $\gamma\simeq 0.01$, based on the work of Giersz \& Heggie
(1997). Second, as $N^\ast$
can become very small, the factor $\ln\gamma N^\ast$ in these
expressions was modified to $\ln(1 + \gamma N^\ast)$.  When $N^\ast =
200$, for example, the value of $dt/dt^\ast$ actually used is too large
by a factor of about $1.6$.  Finally, for astrophysical reasons it was
deemed sufficient to terminate each simulation by the time $t = 20$Gyr.

\bigskip
\noindent
3 Numerical Results
\medskip\noindent
3.1 Tests of the scaling strategy
\smallskip
\noindent
3.1.1 $N$-Dependence of Scaling
\smallskip
In the previous section we have introduced three ways of scaling the
time in a small
$N$-body simulation to that of a typical globular cluster.  We refer to
these respectively as {\sl $t_c$-scaling} (equation 3), {\sl $t_r$-scaling} (equation
5) and {\sl variable scaling} (equation 7).
The most straightforward test of any of these is
the comparison of $N$-body simulations carried out with different values
of $N^\ast$.   This is the purpose of the present subsection.  In \S3.2
we go on to carry out a comparison between the $N$-body results and
those of Fokker-Planck models.  

Fig.2 shows an example of the same model, specified by the information
in the title, computed with different $N^\ast$, and using $t_r$-scaling
(equation 5).
The uppermost curve in
each plot shows the tidal radius, and the remainder are $0.1$\%, $1$\%, $10$\% and
$50$\%  Lagrangian
radii, as defined in the caption.  From a comparison of
the two plots it is gratifying to note that core collapse ends at very
nearly the same time (about $10$Gyr).  Towards the end
of the simulations, however, the larger of the two (Fig.2b, $N^\ast=
8192$)
loses mass at a relatively larger rate.  This can be seen from the
uppermost curve (the tidal radius), as $r_t\propto M^{1/3}$.  This is
an example of a general trend which will be discussed in connection with
Table 1 below.  

Table 1 shows a more extensive set of results, though these models were
computed with variable scaling (equation 7), and each model is now
summarised by only a small number of data.  These data include the time
at which the model either dissipated or was terminated, $t_{max}$,
the time at
the end of core collapse, $t_{cc}$, and the mass and half-mass 
radius  (in $N$-body units) at this time
and at $15$Gyr.  Many of the models, however, do not
reach core collapse, and of these models most also disappear before
$15$Gyr.  

For fixed $W_0$ and $\alpha$, the results from different $N^\ast$ are in
qualitative agreement.  Quantitatively, a systematic result emerges,
which is that the larger model ($N^\ast = 8192$) tends to lose mass somewhat faster than
the smaller model (cf. also Fig.2).  To quantify this we have determined
the time (in Gyr) at which the value of $M^\ast$ in the smaller model
takes the value at $15$Gyr in the larger model.  The resulting five
values range from $15.3$ to $17.4$Gyr, with a mean of $16.6$.  This
gives a quantitative estimate of about 10\% for the order of magnitude
of the discrepancy in the rate of loss of mass.

\smallskip\noindent
3.1.2 Comparison of Different Scalings
\smallskip

Though much was made in \S2.3 about the desirability of variable
scaling, in practice the differences between results produced by the
different scaling methods are not usually very large, at least for initial
particle numbers as large as $N^\ast =8192$.  The basis for this
assessment is the data in Table 2, which compares different types of
scaling for the same clusters.   Results for $t_c$-scaling have been
taken from Fukushige \& Heggie (1995).

Let us first concentrate on those clusters which dissipate promptly,
either during or immediately following the period of heavy mass loss by
stellar evolution.  These are the three models in this table 
for which Fukushige \&
Heggie obtained a definitive result, viz. $W_0 = 3$, $\alpha = 1.5$ and
$2.5$, and $W_0 = 5$, $\alpha = 1.5$.  For such clusters the dissolution
time estimated by $t_r$-scaling is systematically too large.  The results from
variable scaling are better, and we include here the case $W_0 = 3$,
$\alpha = 1.5$, even though the lifetime given by variable scaling is
considerably {\sl less} than that of Fukushige \& Heggie; their Fig. 6
suggests that statistical differences between models will lead to
differences of order $0.03$Gyr, and that the result stated in Table 2
(i.e. $0.11$Gyr) is
near the upper end of the range of values.

Though variable scaling gives a better result for $W_0 = 3$, $\alpha =
2.5$ than $t_r$-scaling, it is still poor.  Inspection of Fig.7 in
Fukushige \& Heggie (1995) suggests that their result is quite robust in
this case.  Though data is not given in Table 2, we have found that, for
smaller particle numbers, the disagreement is still
larger, as expected.

The model that we have been discussing ($W_0 = 3$, $\alpha = 2.5$) 
is expected to be difficult to simulate, because it lies
close to the border between those clusters which dissipate promptly
(without significant relaxation) and those that can survive for several
Gyr.  Another problematic case of this kind is $W_0 = 7$, $\alpha =
1.5$.  Fukushige \& Heggie stopped their calculation at $1.6$Gyr, when the
mass of
this cluster was approximately $M^\ast = 0.05$ (their Fig.1), whereas
our model with variable scaling dissolved at $1.2$Gyr.  It seems likely
that the model of Fukushige \& Heggie, despite the softening they
used, is significantly affected by two-body
relaxation.  This is the longest of their simulations (to $t^\ast =
3000$), and their Fig.8 shows that even a shorter run (to $t^\ast =
1000$, though with different model parameters) 
may be significantly affected by two-body relaxation.
If so, the resulting core evolution could make the cluster too robust to
the effects of mass loss, and so prolong its life.

For the remaining five sets of models presented in Table 2 (which, as it
happens, are
those that survive to $15$Gyr) the results are
more straightforward.  For these models the work of Fukushige \& Heggie gives
only a
lower limit, and can be ignored.  The results from variable- and
$t_r$-scaling are quite consistent, the only noticeable general trend
being that the models with $t_r$-scaling evolve more slowly.  The one
exception to this trend is the core collapse time for $W_0 = 7$, $\alpha
= 2.5$, but this quantity is known to be subject to considerable
statistical fluctuation (Spurzem \& Aarseth 1996). 

\medskip\noindent
3.2 Comparison with Fokker-Planck Results

In this subsection we consider the second main purpose of the
investigation, which is to compare $N$-body results with Fokker-Planck
results for the same cluster parameters, partly as a consistency check,
and partly as a check on the validity of the Fokker-Planck model, which
is likely to remain a standard tool for the investigation
of the dynamical evolution of globular clusters for some time to come.

The basic information we present is given in Tables 3 and 4, which
compare a number of data for our $N$-body models and a {\sl subset} of
the Fokker-Planck models of Chernoff \& Weinberg (1990):  we omit their
models in which $W_0 = 1$ initially.
Note also that our models with $W_0 = 5$ are omitted here
because this was not one of the values which Chernoff \& Weinberg
included. All $N$-body
models use variable scaling (eq.(7)).

A factor to be borne in mind in interpreting the Fokker-Planck data is
that the time $t_{max}$ does not correspond to the time at which the
model loses all its mass.  As Chernoff \& Weinberg explain, there comes
a point at which the method fails to find a new structure corresponding
to the evolving distribution function, and this is interpreted as
indicating the imminent rapid dissolution of the model.

The caption in Table 3 (and those in Table 4, to be discussed below)
refers to one of the 
``families" adopted by
Chernoff \& Weinberg (cf.\S2.1).
For the orbit-averaged technique
used by Chernoff \& Weinberg, the evolution is the same for all members
of one family, given the
initial value of $W_0$ and the initial mass
function.  Thus a low-mass cluster at large galactocentric radius will have the same
predicted evolution as a cluster of high mass at an appropriate small
radius.
Comparison of the models in Table 3 illustrates this
scaling in the $N$-body data, along with a comparison with the
Fokker-Planck data.  In each triplet of models the two $N$-body models
belong to the same family of Chernoff \& Weinberg (family 1), but at
different galactocentric radii.  To avoid complications with the residual
$N$-dependence of
the results of the $N$-body models (cf.\S3.1.1), all models in this Table
share the
same initial value of
$N^\ast$, taken to be $4096$.  
The data of Table 3 shows yet again that the models divide into two types:
those that
dissolve in much less than
a Hubble time, and those in which the $N$-body models survive to at
least $20$Gyr.  

Let us consider first the rapidly dissolving models, i.e. those with
$(W_0,\alpha) = (3,1.5)$, $(3,2.5)$ and $(7,1.5)$.  Though the $N$-body
models indicate systematically longer lifetimes at larger radii, the
evidence is in fact not significant.  We have already argued (in
discussing Table 2) that the difference in the lifetimes of the two
models with $W_0 = 3$, $\alpha = 1.5$, would probably be within
statistical fluctuations even for simulations with $N^\ast = 8192$.  A
similar argument probably applies to the $N$-body models with $W_0 = 7$,
$\alpha = 1.5$, though for somewhat more subtle reasons.  These models
rapidly lose most of their mass, retaining only about 20\% by $0.5$Gyr. 
The subsequent evolution is therefore subject to the sizeable fluctuations
between different $N$-body models with rather small $N^\ast$.

Of the three models which dissolve rapidly, the Fokker-Planck and
$N$-body results are consistent only for the model we have just
discussed, i.e. $W_0 = 7$, $\alpha = 1.5$.  The other two models ($W_0 =
3$, $\alpha = 1.5$ and $2.5$) confirm the finding of Fukushige \& Heggie
(1995) that the Fokker-Planck model underestimates the true lifetime by
a factor of order $10$ in some cases.  As they suggested, a plausible
reason for this is that the
Fokker-Planck models necessarily assume that $t_M$, the time scale on which the
cluster loses mass, much exceeds
$t_{cr}$, and that the lifetime is underestimated when this assumption
is not valid.

Now we turn to the more straightforward comparison of the three models
in Table 3 which survive for at least a Hubble time.  The $N$-body
models illustrate rather well the fact that the evolution is similar for
clusters within the same family but at different $R_g$.  The results
also agree rather well with the Fokker-Planck data.  Though it might
seem that the Fokker-Planck models evolve somewhat faster than these
$N$-body models, it must be recalled (cf.  discussion of Table 1) that
larger $N$-body models would also do so.

Table 4 is the main vehicle for examining the relation between the
$N$-body data and the Fokker-Planck data.  Here all models are placed at
about $4$kpc, as the evidence of Table 3 suggests that the results will
be independent of $R_g$ within a given family, at least for long-lasting
models.   In fact we have omitted those
models that dissolve rapidly, as an adequate comparison has been
provided by Fukushige \& Heggie.

The data on core collapse in this Table is rather meagre and adds little
to what was learned from Table 3, except that the models for Family 1
have larger $N^\ast$.  It confirms that there appear to be no systematic
differences in either the time or the total mass at core collapse. 

Now
let us consider the mass at 15Gyr. The absence of this data in two of the
Fokker-Planck models of Family 1 is due to the fact that Chernoff \&
Weinberg stopped their calculations at the time of core collapse, which
occurs before $15$Gyr for these models.  For the one remaining model in
this family, and all models in the other families, 
there is evidence of some systematic differences in the
mass at $15$Gyr.  In particular, the disagreement is most pronounced for the
models with $W_0 = 3$ and $\alpha = 3.5$, especially for the first two
families, i.e. those in which two-body relaxation is fastest.  The
Fokker-Planck models yield substantially smaller masses at $15$Gyr.  If we assume
that the mass decreases approximately linearly with time, the differences
correspond to differences in the lifetime which are in the range of
$13$--$23$\%.  The comparisons in Table 2, however, suggest that
the disagreement is well within the residual $N^\ast$-dependence of the
$N$-body results, and has the expected sign:  the mass at $15$Gyr
decreases slightly with increasing $N^\ast$, 
and the Fokker-Planck results
are also smaller than the $N$-body results.  These remarks apply to the
models with $W_0 = 3$ and $\alpha = 3.5$; for other models the 
discrepancy in the mass at $15$Gyr is much more
modest.

\vfill\eject
\noindent
4. Discussion and Conclusions
\medskip\noindent 
4.1 Anisotropy
\smallskip
Though the main purposes of this paper are to investigate the scaling of
$N$-body models and to compare results with Fokker-Planck data, it is a
pity to leave the $N$-body models without drawing attention to some
other interesting results.  One of these is the evolution of the mass
function and the proportion of degenerate stars, and we refer
interested readers to Vesperini \& Heggie (1997), where 
detailed and comprehensive results are presented (though with a more
realistic lower limit to the initial mass function, and disk shocking).

Here we concentrate on the anisotropy of the models, which Chernoff \& Weinberg
did not address, as they used an isotropic Fokker-Planck model.  For the
purposes of this discussion we
adopt the usual parameter $\beta = 1 - \langle v_t^2\rangle/\langle
v_r^2\rangle$, where $v_t^2$ is half of the square of the transverse
component of velocity of one star and $v_r$ is the radial component,
i.e. in the direction away from the density centre.  Only stars inside the
tidal radius are included in the averaging, which is not mass-weighted.
Initially there is a short period in which $\beta > 0$, and this is attributable
to the radial outward drift of stars during the early phase of rapid
mass loss.   Thereafter the sense of the anisotropy reverses, i.e.
$\beta <0$, and there is a {\sl slight} predominance of transverse
velocities.  Nevertheless the anisotropy is small, values of order
$\beta\simeq-0.1$ being typical.  

In an {\sl isolated} model, $\beta$ would become positive (e.g. Giersz
\& Heggie 1996), and the
negative values in our model are attributable to the tidal field.  This
has two effects:  in the first place it alters the angular momentum of
the stars, as it is not spherically symmetric (cf.eq.[1]); and, secondly,
it preferentially removes stars on nearly radial orbits, as these are
most likely to escape.   We attribute the negative values of $\beta$ to
the second effect mainly, as similar negative values of $\beta$ are
observed in models (not otherwise  discussed in this paper) in which the
tidal field of eq.(1) is replaced by a spherical {\sl cutoff}, as in most
implementations of the Fokker-Planck model. 

The smallness of the values of $\vert\beta\vert$ helps
to justify the use of an isotropic model such as the Fokker-Planck
model of Chernoff \& Weinberg.  On the other hand, our result also 
has a variety of
implications for the interpretation of 
observational data.  There are a few clusters for which careful study of
proper motions requires the presence of significant radial anisotropy,
i.e. $\beta > 0$ (e.g. Cudworth 1979 for the cluster M3).  There is a
larger sample of clusters where careful dynamical modelling (using
multi-mass anisotropic King-Michie models) again requires the presence
of radial anisotropy (e.g. Meylan \& Mayor 1991).  Here the requirement
of substantial anisotropy is less direct, as the models which are fitted
are constrained by theoretical considerations (e.g. the choice of
distribution function), and in principle it is possible that an incorrect
choice of distribution function is compensated by the
anisotropy when
fitting to the observational data.  Nevertheless, if the evidence
for anisotropy is accepted, the contradiction with the anisotropy found
in our models may only indicate that the initial conditions adopted in
our survey differ too widely from those appropriate to the observed
clusters.  For example, all our models are initially tidally limited:
the edge of the model is placed initially at the tidal radius.  It is
possible that initially much more compact models would develop strongly
positive anisotropy.  They would behave much more like isolated models
until, after sufficient expansion in the post-collapse phase of
evolution, their development might more closely resemble that of the
models we have studied.  Interestingly, it was concluded by 
Phinney (1993), on
quite different grounds, that only an initially very condensed cluster
would evolve into an object resembling the cluster M15 at the present
day.
\medskip\noindent
4.2 Conclusions
\smallskip

Now we close with a summary of the main conclusions of this paper.  It
has been concerned with the dynamical evolution, over a time up to
$20$Gyr, of a set of models for the globular star clusters of the
Galaxy.  The evolutionary mechanisms considered here include all
gravitational interactions between the members of a cluster, the tidal
field of the Galaxy, and instantaneous mass loss at the end of main
sequence evolution.  We do not include the effects of finite stellar
radii and collisions, or primordial binaries.  The initial mass function
is a power law between masses of $15$ and $0.4M_\odot$, with index
$\alpha$. The initial model is a King model, whose concentration is
specified by the dimensionless central potential $W_0$.

In qualitative agreement with the results of Chernoff \& Weinberg (1990)
we find that the fate of the clusters divides them into two classes: 
those that disrupt relatively rapidly (in at most one or two Gyr) as a result of
mass loss by stellar evolution and tidal stripping, without being
significantly affected by two-body relaxation; and those that survive
this initial phase of heavy mass loss and evolve towards core collapse. 
For clusters of initial mass about $1.5\times10^5M_\odot$ and
galactocentric radius of about $4$kpc, clusters which
disrupt are those with $W_0 = 3$ and $\alpha\ltorder2.5$, and those with
$W_0 = 5$ or $7$ and
$\alpha\ltorder1.5$.  In general those with larger $W_0$ and/or larger $\alpha$
enter the phase in which their evolution is dominated by relaxation. 

For those clusters which disrupt promptly, a comparison between $N$-body
and Fokker-Planck models was carried out by Fukushige \& Heggie (1995).  In
this paper we have concentrated on the longer-lived systems which become
relaxation-dominated.  For some parameter values the Fokker-Planck and
$N$-body results agree remarkably well;  the time of core collapse of
models with $W_0= 7$ and $\alpha\gtorder2.5$ agree as well as could be
expected, bearing in mind the fact that there will be different collapse
times even for different
$N$-body realisations with the same initial parameters.   More
systematic differences, of order $10$\%, 
arise for models which lie closer to the
borderline between early disruption and relaxation-dominated evolution,
e.g. the model with $W_0=3$ and $\alpha=3.5$.  For
these models several lines of evidence (e.g. the mass at the present
day, taken here as $15$Gyr) imply that the Fokker-Planck models evolve
somewhat more rapidly than the $N$-body models.  
The sense of the disagreement is consistent with the conclusion of Fukushige \& Heggie
for the clusters which disrupt quickly, though the size of the
disagreement is much smaller for the clusters which do not dissolve
rapidly.  Indeed the residual $N^\ast$-dependence of the rate of escape
from the $N$-body models 
may explain part or all of the discrepancy.

Besides a comparison with Fokker-Planck models, the other main purpose
of this paper was an investigation of the way in which the evolution of 
small $N$-body
models may be scaled (in time) to that of globular clusters, where the
number of stars is much larger.  We have considered three possible
scalings: (i) a scaling which ensures that the crossing time of the
$N$-body model scales correctly to that of a real cluster; (ii) a
scaling which ensures that the model relaxes at the same rate as the
real cluster; and (iii) a hybrid, or variable, scaling, which progresses
from the first type of scaling to the second as the mass loss from
stellar evolution slows down.

The first type of scaling was applied by Fukushige \& Heggie.  It
successfully allows the scaling of $N$-body data for clusters which
dissipate promptly without significant relaxation.  The model {\sl relaxes}
too quickly, however, and so the scaling of the simulation is no longer
valid as soon as relaxation in the model becomes significant.   The
second type of scaling produces results which appear to be correct for
clusters which survive the initial phase of mass loss and enter the
relaxation dominated phase of evolution.  Unless the number of particles
in the simulation is quite large, however, it can produce misleading
results for those clusters which dissipate promptly.

The third (variable) scaling was devised to incorporate the
advantages of both fixed types of scaling.  It produces qualitatively
correct results for both kinds of clusters, i.e. those that dissipate
rapidly and those that become relaxation-dominated.  For clusters of the
latter type its quantitative
results are indistinguishable from those obtained by scaling of type
(ii).  It is the most successful scaling method for clusters on the
borderline between rapid destruction and long-term survival, e.g. King
models with $W_0 = 7$ and $\alpha = 1.5$ (for our chosen initial range
of stellar masses).

We recommend variable scaling as a means for using small $N$-body
simulations to model the evolution of globular star clusters.  Its
advantage is that it can be used (with some care) to model the evolution
of clusters which dissipate promptly as well as those with long
lifetimes.  Though this might not seem an important property for the
galactic globular cluster system, it is relevant if $N$-body models are
to be used for studying the evolution of other globular cluster systems, as
in the Magellanic Clouds,  which contain both young and old clusters.

The results obtained in this paper are based on models which
incorporate only a subset of the many dynamical and evolutionary processes which are
needed in a full $N$-body description of a globular cluster (e.g.
Aarseth 1996), but the problem of how to scale models cannot be avoided as
long as the number of particles in a feasible $N$-body simulation is
much smaller than in a real globular star cluster.  In a further paper
(Aarseth \& Heggie 1998) 
we shall consider how the radii of the stars may be scaled in order to
simulate stellar collisions.

\bigskip\noindent
Acknowledgments.  This work has been supported by SERC/PPARC under grant
number GR/J79461, which funded the GRAPE-4 hardware.  
We are deeply indebted to several members of the
GRAPE group at the University of Tokyo for their unstinting assistance
in the planning, provision and maintenance of this equipment, especially
Prof. D. Sugimoto and Drs J. Makino and M. Taiji.  We thank Dr S.
Sigurdsson for helping to run the hardware, and Dr E. Vesperini for much
help with software and scientific aspects of the project.  Finally we
thank the referee, who suggested a number of changes which helped shape
the final form of the paper.

\bigskip
\noindent
References
\medskip

{\parindent=-0.5truein
\leftskip=0.5truein

Aarseth S.J., 1985, in Brackbill J.U., Cohen B.I., eds, Multiple 
Time Scales.  Academic Press, New York, p.377

Aarseth S.J., 1996, in Hut P., Makino J., eds, Dynamical Evolution of
Star Clusters, IAU Symp. 174. Kluwer, Dordrecht, p.161

Aarseth S.J., Heggie D.C., 1998, in preparation

Ahmad A., Cohen L., 1973, J. Comp. Phys., 12, 389

Applegate J.H., 1986, ApJ, 301, 132

Chandrasekhar S., 1942, Principles of Stellar Dynamics. Univ. of Chicago
Press, Chicago; also Dover Pub. Inc., New York, 1960

Chernoff D.F., Shapiro S.L., 1987, ApJ, 322, 113

Chernoff D.F., Weinberg M.D., 1990, ApJ, 351, 121

Cudworth K.M., 1979, AJ, 84, 1312

Drukier G.A., 1996, MNRAS, 280, 498

Einsel C.R.W., Spurzem R., 1998, MNRAS, submitted

Fukushige T., Heggie D.C., 1995, MNRAS, 276, 206

Gao B., Goodman J., Cohn H., Murphy B., 1991, ApJ, 370, 567

Giersz M., Heggie D.C., 1996, MNRAS, 279, 1037

Giersz M., Heggie D.C., 1997, MNRAS, 286, 709

Goodman J., 1983a, PhD Thesis, Princeton Univ.

Goodman, J., 1983b, ApJ, 270, 700

Goodman J.,  Hut P., 1989, Nature, 339, 40

Goodwin S.P., 1997, MNRAS, 284, 785

Heggie D.C., Aarseth S.J., 1992, MNRAS, 257, 513

Heggie D.C.,  Mathieu R.D., 1986,  in  Hut P.,   McMillan S.L.W., 
eds, The Use of Supercomputers
in Stellar Dynamics. Springer-Verlag, Berlin, p.233

Heggie D.C., Ramamani N.,  1995, MNRAS, 272, 317

Hills J.G., 1980, ApJ, 235, 986

Hut P., McMillan S., Goodman J., Mateo M., Phinney E.S., Pryor C., Richer H.B., Verbunt F., Weinberg M., 1992,  PASP, 104, 981

King I.R., 1966, AJ, 71, 64

Lada C.J., Margulis M., Dearborn D.S., 1984, ApJ, 285, 141

Makino J., 1996, ApJ, 471, 796

Makino J.,  Aarseth S.J., 1992, PASJ, 44, 141

Makino J., Kokubo E., Taiji M., 1993, PASJ, 45, 349

Makino J., Taiji M., Ebisuzaki T., Sugimoto D., 1997, ApJ, 480, 432

Meylan G., Mayor M., 1991, A\&A, 250, 113

Mikkola S., Aarseth S.J., 1993, Celes. Mech. Dyn. Astron., 57, 439

Mikkola S., Aarseth S.J., 1996, Celes. Mech. Dyn. Astron., 64, 197

Milone E.F., Mermilliod J.-C., 1996, eds, The Origins, Evolution and
Destinies of Binary Stars in Clusters, ASP Conf. Ser. 90. ASP, San
Francisco.

Phinney E.S., 1993 in Djorgovski S.G., Meylan G., eds, Structure and
Dynamics of Globular Clusters, ASP Conf. Ser. 50.  ASP, San 
Francisco, p.141

Spitzer L., Jr., 1987, Dynamical Evolution of Globular Clusters. 
Princeton Univ. Press, Princeton.

Spitzer L., Jr., Hart M.H., 1971, ApJ, 164, 399

Spurzem R., Aarseth S.J., 1996, MNRAS, 282, 19

Takahashi K., 1995, PASJ, 47, 561

Vesperini E., Chernoff D.F., 1994, ApJ, 431, 231

Vesperini E., Heggie D.C., 1997, MNRAS, 289, 898

}

\bigskip

\vfill\eject
\baselineskip=\normalbaselineskip
\centerline{Table 1}
\medskip
\centerline{$N$-Dependence of Models with Variable Scaling}
\medskip
\settabs 12 \columns
\+$W_0$	&$\alpha$	&$R_g$	&$N^\ast$	&$t_{cpu}$	&$t^\ast_{max}$	&$t_{max}$	&$t_{cc}$	&$M^\ast_{cc}$	&$r^\ast_{h,cc}$	&$M^\ast_{15}$	&$r^\ast_{h,15}$	\cr
\+	&		&kpc	&	&hr		&
&Gyr		&Gyr		&		&		\cr
\+$3$	&$1.5$	 	&$3.7$	&$4096$	&$0.3$		&$39$	&$0.075$		&--		&--		&--	&--	&--		\cr
\+$3$	&$1.5$	 	&$3.7$	&$8192$	&$0.8$		&$37$		&$0.071$		&--		&--		&--		&--	&--		\cr
\smallskip
\+$3$	&$2.5$	 	&$4.0$	&$4096$	&$1.0$		&$169$	&$3.0$		&--		&--		&--	&--	&--		\cr
\+$3$	&$2.5$	 	&$4.0$	&$8192$	&$3.3$		&$215$	&$2.1$		&--		&--		&--	&--	&--		\cr
\smallskip
\+$3$	&$3.5$	 	&$4.1$	&$4096$	&$3.3$		&$432$		&$20$		&--		&--		&--		&$0.390$	&$0.69$		\cr
\+$3$	&$3.5$		&$4.1$	&$8192$	&$12$		&$669$		&$18.5^+$		&--		&--		&--		&$0.355$	&$0.67$		\cr
\smallskip
\+$5$	&$1.5$		&$3.7$	&$4096$	&$0.6$		&$123$		&$0.29$		&--		&--	&--		&--	&--		\cr
\+$5$	&$1.5$		&$3.7$	&$8192$	&$1.7$		&$144$		&$0.23$		&--		&--	&--		&--	&--		\cr
\smallskip
\+$5$	&$2.5$		&$4.0$	&$4096$	&$12^{[4]}$	&$1161$		&$16.7$		&$13$		&$0.086$	&$0.48$		&$0.0385$	&$0.42$		\cr
\+$5$	&$2.5$ 		&$4.0$	&$8192$	&$13$		&$1789$		&$16.6$		&$13.5$		&$0.067$	&$0.40$		&$0.032$	&$0.32$		\cr
\smallskip
\+$5$	&$3.5$		&$4.1$ 	&$4096$	&$7.0$		&$675$		&$20$		&--		&--		&--		&$0.515$	&$0.69$		\cr
\+$5$	&$3.5$		&$4.1$	&$8192$ &$\simeq27$	&$1130$		&$20$		&--		&--		&--		&$0.477$	&$0.67$		\cr
\smallskip
\+$7$	&$1.5$		&$3.7$	&$4096$ &$1.3$		&$438$		&$1.2$		&--		&--		&--		&--		&--		\cr
\+$7$	&$1.5$		&$3.7$ 	&$8192$ &$4.2$		&$658$		&$1.2$		&--		&--		&--		&--		&--		\cr
\smallskip
\+$7$	&$2.5$		&$4.0$	&$4096$	&$10$		&$2351$		&$20$		&$13$		&$0.21$		&$0.83$		&$0.148$	&$0.76$		\cr
\+$7$	&$2.5$		&$4.0$	&$8192$	&$33$		&$3825$		&$18.7$		&$11$		&$0.210$	&$0.77$		&$0.089$	&$0.60$		\cr
\smallskip
\+$7$	&$3.5$		&$4.1$	&$4096$	&$22$		&$1368$		&$20$		&$10.4$		&$0.606$	&$0.94$		&$0.476$	&$0.92$		\cr
\+$7$	&$3.5$		&$4.1$	&$8192$	&$85$		&$2277$		&$20$		&$9.2$		&$0.617$	&$0.89$		&$0.435$	&$0.94$		\cr
\medskip
{\parindent = 0pt

Notes

1.  An asterisk in the top line denotes a quantity in $N$-body units in
an $N$-body model.  The values of $W_0$, $\alpha$ and $N^\ast$ are
initial values. Quantities at core collapse  are denoted by the
subscript cc, and at $15$Gyr by a subscript $15$.   The half-mass radius
(of all particles inside the tidal radius) is denoted by $r_h$.  

2. The simulations end at $t_{max}$; where this value is less than 20,
the calculation stopped with less than $10$ bound particles, except that
a plus indicates that it ended with a software or hardware error.

3. In all these models the initial mass is $M=1.49\times10^5M_\odot$. 

4. This model was mainly computed without the HARP board.

}
\vfill\eject
\centerline{Table 2}
\medskip
\centerline{Comparison of Different Time Scalings}
\medskip
\settabs 10\columns
\+$W_0$	&$\alpha$	&$R_g$	&Scaling	&$t_{max}$	&$t_{cc}$	&$M^\ast_{cc}$	&$r_{h,cc}^\ast$&$M^\ast_{15}$	&$r_{h,15}^\ast$\cr
\+	&		&kpc	&		&Gyr		&Gyr		&		&		&		&		\cr
\smallskip
\+$3$	&$1.5$		&$3.7$	&$t_c$		&$0.11$	&--		&--		&--		&--	&--		\cr
\+$3$	&$1.5$		&$3.7$	&var.		&$0.071$	&--		&--		&--		&--	&--		\cr
\+$3$	&$1.5$		&$3.7$	&$t_r$		&$0.14$	&--		&--		&--		&--	&--		\cr
\smallskip
\+$3$	&$2.5$		&$4.0$	&$t_c$		&$0.91$	&--		&--		&--		&--	&--		\cr
\+$3$	&$2.5$		&$4.0$	&var.		&$2.1$	&--		&--		&--		&--	&--		\cr
\+$3$	&$2.5$		&$4.0$	&$t_r$		&$2.9$	&--		&--		&--		&--	&--		\cr
\smallskip
\+$3$	&$3.5$		&$4.1$	&$t_c$		&$>3.9$		&--		&--		&--		&		&--		\cr
\+$3$	&$3.5$		&$4.1$	&var.		&$18.5^+$	&--		&--		&--		&$0.355$	&$0.67$		\cr
\+$3$	&$3.5$		&$4.1$	&$t_r$		&$19^+$		&--		&--		&--		&$0.370$	&$0.67$		\cr
\smallskip
\+$5$	&$1.5$		&$3.7$	&$t_c$		&$0.24$	&--		&--		&--		&--	&--		\cr
\+$5$	&$1.5$		&$3.7$	&var.		&$0.23$	&--		&--		&--		&--	&--		\cr
\+$5$	&$1.5$		&$3.7$	&$t_r$		&$0.42$	&--		&--		&--		&--	&--		\cr
\smallskip
\+$5$	&$2.5$		&$4.0$	&$t_c$		&$>2.3$		&--		&--		&--		&--	&--		\cr
\+$5$	&$2.5$		&$4.0$	&var.		&$16.6$	&$13.5$		&$0.067$	&$0.40$		&$0.032$	&$0.32$		\cr
\+$5$	&$2.5$		&$4.0$	&$t_r$		&$18.6$	&$15$		&$0.0847$	&$0.42$		&$0.0847$	&$0.42$		\cr
\smallskip
\+$5$	&$3.5$		&$4.1$	&$t_c$		&$>2.4$		&--		&--		&--		&--		&--		\cr
\+$5$	&$3.5$		&$4.1$	&var.		&$20$		&--		&--		&--		&$0.477$	&$0.67$		\cr
\+$5$	&$3.5$		&$4.1$	&$t_r$		&$20$		&--		&--		&--		&$0.480$	&$0.65$		\cr
\smallskip
\+$7$	&$1.5$		&$3.7$	&$t_c$		&$>1.6$		&--		&--		&--		&--		&--		\cr
\+$7$	&$1.5$		&$3.7$	&var.		&$1.2$	&--		&--		&--		&--		&--		\cr
\+$7$	&$1.5$		&$3.7$	&$t_r$		&$4.8$	&$4.0$		&$0.0069$	&$0.30$		&--		&--		\cr
\smallskip
\+$7$	&$2.5$		&$4.0$	&$t_c$		&$>1.1$		&--		&--		&--		&--		&--		\cr
\+$7$	&$2.5$		&$4.0$	&var.		&$18.7$	&$11$		&$0.210$	&$0.77$		&$0.089$	&$0.60$		\cr
\+$7$	&$2.5$		&$4.0$	&$t_r$		&$20$		&$10.0$		&$0.2666$	&$0.75$		&$0.1113$	&$0.67$		\cr
\smallskip
\+$7$	&$3.5$		&$4.1$	&$t_c$		&$>1.2$		&--		&--		&--		&--		&--		\cr
\+$7$	&$3.5$		&$4.1$	&var.		&$20$		&$9.2$		&$0.617$	&$0.89$		&$0.435$	&$0.94$		\cr
\+$7$	&$3.5$		&$4.1$	&$t_r$		&$20$		&$10$		&$0.594$	&$0.89$		&$0.438$	&$0.89$		\cr

\medskip

{\parindent = 0pt

Notes

1.  Most of the notation is defined in the notes to Table 1.  

2. Under the heading ``Scaling", $t_c$ denotes scaling by the crossing
time, eq.(3); $t_r$ denotes scaling by the relaxation time, eq.(5); and ``var."
denotes variable scaling, eq.(7).  Results for $t_c$-scaling are taken from
Fukushige \& Heggie (1995), and for several models the lifetime is only
a lower limit.

3. In all these models the initial mass is $M=1.49\times10^5M_\odot$.   All runs had $N^\ast = 8192$ particles initially.

}
\vfill\eject
\centerline{Table 3}
\medskip
\centerline{Comparison of Fokker-Planck and $N$-body results for
Family1}
\medskip
\settabs 7\columns
\+$W_0$	&$\alpha$	&Model	&$t_{cc}$	&$t_{max}$	&$M^\ast_{cc}$	&$M^\ast_{15}$	\cr
\smallskip
\+	&		&	&Gyr		&Gyr		&		&		\cr
\smallskip         
\smallskip
\+$3$	&$1.5$		&$3.7$	&--		&$0.075$	&--		&--		\cr
\+$3$	&$1.5$		&$9.2$	&--		&$0.12$		&--		&--		\cr
\+$3$	&$1.5$		&FP	&--		&$0.014$	&--		&--		\cr
\smallskip
\+$3$	&$2.5$		&$4.0$	&--		&$3.0$		&--		&--		\cr
\+$3$	&$2.5$		&$10.0$	&--		&$3.4$		&--		&--		\cr
\+$3$	&$2.5$		&FP	&--		&$0.28$		&--		&--		\cr
\smallskip
\+$3$	&$3.5$		&$4.1$	&--		&$20$		&--		&$0.390$	\cr
\+$3$	&$3.5$		&$10.4$	&--		&$20$		&--		&$0.419$	\cr
\+$3$	&$3.5$		&FP	&$21.5$		&--		&$0.078$	&$0.23$		\cr
\smallskip
\+$7$	&$1.5$		&$3.7$	&--		&$1.2$		&--		&--		\cr
\+$7$	&$1.5$		&$9.2$	&--		&$3.2$		&--		&--		\cr
\+$7$	&$1.5$		&FP	&--		&$1.0$		&--		&--		\cr
\smallskip
\+$7$	&$2.5$		&$4.0$	&$13$		&$20$		&$0.21$		&$0.148$	\cr
\+$7$	&$2.5$		&$10.0$	&$11$		&$20$		&$0.257$	&$0.161$	\cr
\+$7$	&$2.5$		&FP	&$9.6$		&--		&$0.26$		&--		\cr
\smallskip
\+$7$	&$3.5$		&$4.1$	&$10.4$		&$20$		&$0.606$	&$0.476$	\cr
\+$7$	&$3.5$		&$10.4$	&$10.5$		&$20$		&$0.62$		&$0.494$	\cr
\+$7$	&$3.5$		&FP	&$10.5$		&--		&$0.57$		&--		\cr
\medskip
{\parindent = 0pt

Notes

1.  Most of the notation is defined in the notes to Table 1.

2. Under the heading ``Model", a number indicates results of an $N$-body
model (with $N^\ast = 4096$ initially) at the stated galactocentric
distance (in kpc).  ``FP" denotes a Fokker-Planck model of Chernoff \&
Weinberg (1990).  Their models were stopped at the point of dissolution
of the cluster (in which case we give the corresponding time as
$t_{max}$) or at the end of core collapse (at time $t_{cc}$), whichever
is earlier.

3. The initial cluster mass is $5.45\times10^4M_\odot$ at around
$10$kpc, and $1.49\times10^5M_\odot$ at around $4$kpc.  All $N$-body 
models use
variable scaling (eq.[7]).

}

\vfill\eject
\centerline{Table 4}
\medskip
\centerline{Comparison with Fokker-Planck Results}
\medskip
\settabs 7\columns
\+$W_0$	&$\alpha$	&$R_g$ (kpc)&Model&$t_{cc}$ (Gyr)&$M^\ast_{cc}$ &$M^\ast_{15}$\cr
\smallskip         
\centerline{Family 1: $M(0) = 1.49\times10^5M_\odot$}
\smallskip
\+$3$	&$3.5$		&$4.1$	&$8192$	&$>18.5$	&--		&$0.355$	\cr
\+$3$	&$3.5$		&$4.1$	&FP	&$21.5$		&$0.078$	&$0.23$		\cr
\smallskip
\+$7$	&$1.5$		&$3.7$	&$8192$	&($1.2$D)	&--		&--		\cr
\+$7$	&$1.5$		&$3.7$	&FP	&($1.0$D)	&($0.022)$	&--		\cr
\smallskip
\+$7$	&$2.5$		&$4.0$	&$8192$	&$11$		&$0.210$	&$0.089$	\cr
\+$7$	&$2.5$		&$4.0$	&FP	&$9.6$		&$0.26$		&--		\cr
\smallskip
\+$7$	&$3.5$		&$4.1$	&$8192$	&$9.2$		&$0.617$	&$0.435$	\cr
\+$7$	&$3.5$		&$4.1$	&FP	&$10.5$		&$0.57$		&--		\cr
\smallskip
\centerline{Family 2: $M(0) = 4.33\times10^5M_\odot$}
\smallskip
\+$3$	&$3.5$		&$4.1$	&$4096$	&$>20$		&--		&$0.589$	\cr
\+$3$	&$3.5$		&$4.1$	&FP	&$44.4$		&$0.035$	&$0.48$		\cr
\smallskip
\+$7$	&$1.5$		&$3.7$	&$4096$	&($2.0$D)	&--		&--		\cr
\+$7$	&$1.5$		&$3.7$	&FP	&($3.0$D)	&($0.0033)$	&--		\cr
\smallskip
\+$7$	&$2.5$		&$4.0$	&$4096$	&$>20$		&--		&$0.331$	\cr
\+$7$	&$2.5$		&$4.0$	&FP	&$22.5$		&$0.26$		&$0.35$		\cr
\smallskip
\+$7$	&$3.5$		&$4.1$	&$4096$	&$>20$		&--		&$0.698$	\cr
\+$7$	&$3.5$		&$4.1$	&FP	&$31.1$		&$0.51$		&$0.68$		\cr
\smallskip
\centerline{Family 3: $M(0) = 7.65\times10^5M_\odot$}
\smallskip
\+$3$	&$3.5$		&$4.1$	&$4096$	&$>20$		&--		&$0.633$	\cr
\+$3$	&$3.5$		&$4.1$	&FP	&($42.3$D)	&($0.085$)	&$0.58$		\cr
\smallskip
\+$7$	&$1.5$		&$3.7$	&$4096$	&($3.1$D)	&--		&--		\cr
\+$7$	&$1.5$		&$3.7$	&FP	&($4.2$D)	&($0.0080)$	&--		\cr
\smallskip
\+$7$	&$2.5$		&$4.0$	&$4096$	&$>20$		&--		&$0.374$	\cr
\+$7$	&$2.5$		&$4.0$	&FP	&$35.5$		&$0.26$		&$0.40$		\cr
\smallskip
\+$7$	&$3.5$		&$4.1$	&$4096$	&$>20$		&--		&$0.741$	\cr
\+$7$	&$3.5$		&$4.1$	&FP	&$51.3$		&$0.48$		&$0.71$		\cr
\smallskip
\centerline{Family 4: $M(0) = 2.17\times10^6M_\odot$}
\smallskip
\+$3$	&$3.5$		&$4.1$	&$4096$	&$>20$		&--		&$0.692$	\cr
\+$3$	&$3.5$		&$4.1$	&FP	&($43.5$D)	&$(0.28)$	&$0.64$		\cr
\smallskip
\+$7$	&$1.5$		&$3.7$	&$4096$	&($5.0$D)	&--		&--		\cr
\+$7$	&$1.5$		&$3.7$	&FP	&($5.9$D)	&($0.023)$	&--		\cr
\smallskip
\+$7$	&$2.5$		&$4.0$	&$4096$	&$>20$		&--		&$0.443$	\cr
\+$7$	&$2.5$		&$4.0$	&FP	&$83.1$		&$0.25$		&$0.45$		\cr
\smallskip
\+$7$	&$3.5$		&$4.1$	&$4096$	&$>20$		&--		&$0.783$	\cr
\+$7$	&$3.5$		&$4.1$	&FP	&$131.3$	&$0.49$		&$0.77$		\cr
\bigskip

{\parindent=0pt

Notes 

1.  Most of the notation is explained in the notes to the previous
tables.

2.  A number in the column ``Model" indicates that the results are
for an $N$-body model with the initial value of $N^\ast$ as stated.

3.  Though the time of core collapse, $t_{cc}$, is given as ``$>20$" for
those models which stopped at $20$Gyr without collapsing, it is not
possible to state whether these models would reach core collapse before
dissipating.  Where the value in this column is enclosed in brackets,
with a ``D", the model dissipated at the time stated, without previously
undergoing core collapse; for Fokker-Planck models the mass at this
time is entered in parentheses in the next column.

\vfill\eject
\noindent
{\bf Figure Captions}
\bigskip
Fig.1  Time scale of mass loss (eq.(8)) for mass functions with minimum
mass $0.4 M_\odot$ and mass function index $\alpha = 1.5,$ $2.5$ and
$3.5$ (where Salpeter is $2.35$).  The piecewise linear nature of the
curves arises from low-order interpolation in the table of evolution
time as a function of initial stellar mass (Chernoff \& Weinberg 1990).

\medskip
Fig.2 Comparison of results for the same initial and boundary
conditions, but with different initial values of $N^\ast$: (a) $N^\ast = 4096$, (b)
$N^\ast = 8192$.  The tidal radius and four 
Lagrangian radii (corresponding to the innermost $0.1$\%, $1$\%, $10$\% and
$50$\% of the mass within the tidal radius, measured from the density centre)
are
plotted against time.  

}

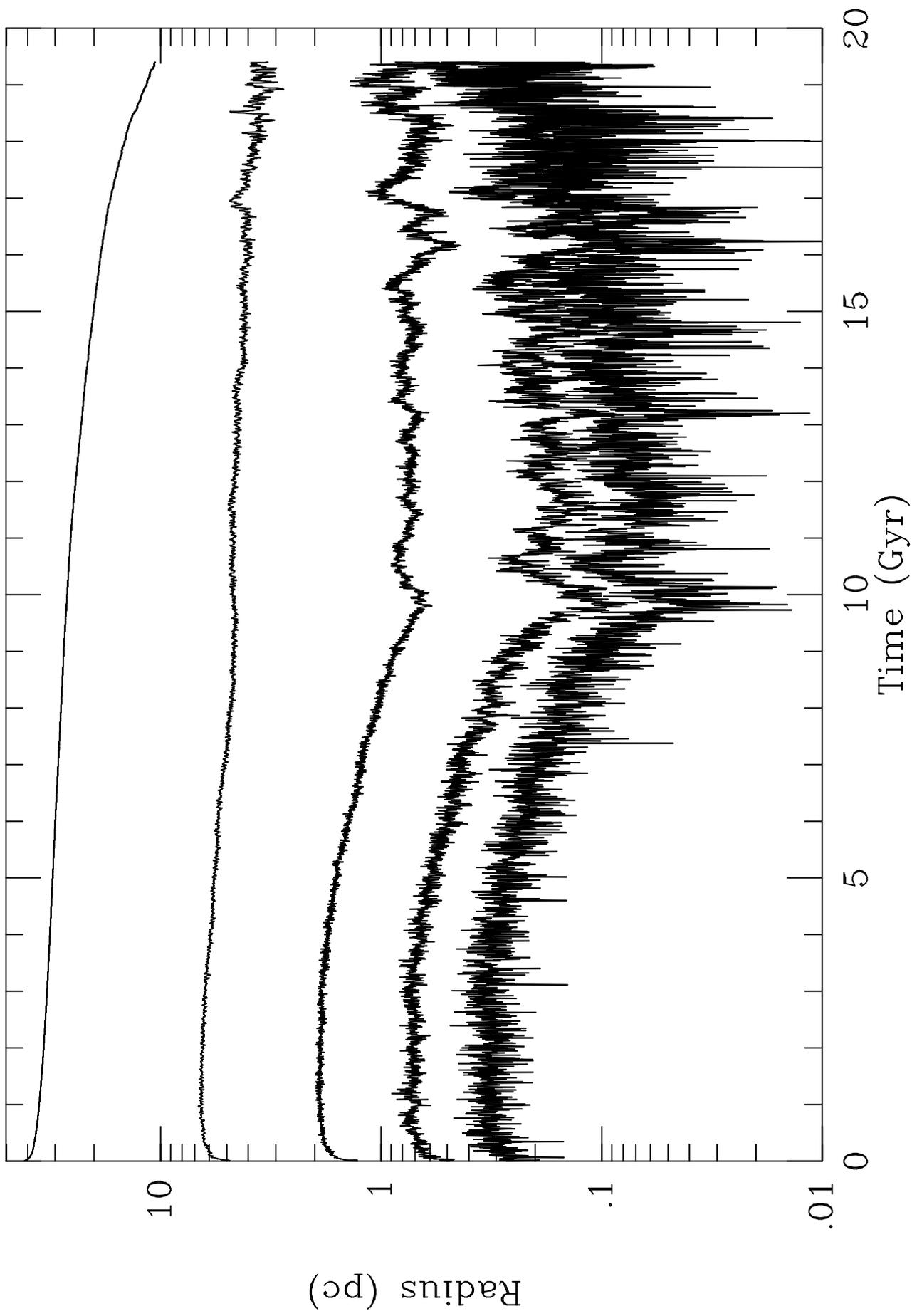
\bye